\documentstyle[12pt]{article}

\begin{document}
\thispagestyle{empty}
\begin{center}

\null
\vskip-1truecm
\vskip2truecm
{\bf WHAT IF $w < -1$ ?\\}
\vskip2truecm
Brett McInnes
\vskip2truecm

Department of Mathematics, National University of Singapore, 10 Kent Ridge Crescent,
Singapore 119260, Republic of Singapore.\\ 
E-mail: matmcinn@nus.edu.sg\\    

\end{center}
\vskip1truecm
\centerline{ABSTRACT}
\baselineskip=15pt

The cosmological ``equation of state parameter", $w$, is equal to $-1$ for a true cosmological constant, and is greater than $-1$ for quintessence. There is a widespread reluctance to consider the remaining possibility, $w < -1$, though in fact there is a variety of theoretical and observational evidence in favour of this range. We briefly discuss some of the objections to $w < -1$ and show that each of them can readily be circumvented. We also briefly consider what string theory has to say about this.

\addtocounter{section}{1}
\section*{1. The Fear of $w < -1$: Horror Vacui Redux}
There is now strong evidence [see for example \cite{kn:sievers}] that our Universe is pervaded by some kind of ``dark energy" with negative pressure. The real nature of this substance is quite unknown, and there is no justification whatever for assuming that it resembles known forms of matter or energy. One should therefore be very cautious about imposing on it one's prejudices as to how respectable forms of matter ought to behave.
 
At this point, essentially all available evidence is consistent with the possibility that $w$, the ratio of the dark energy pressure to its energy density, is equal to $-1$, as predicted by a true cosmological constant. Obviously, therefore, all available evidence is consistent with the possibility that $w$ is slightly {\em less} than $-1$. This fact was first emphasised by Caldwell \cite{kn:caldwell}. However, the idea appears to have encountered opposition, usually expressed covertly by brutal amputations of the data contours in the $w-\Omega_m$ plane. A gratifying exception is provided by \cite{kn:hannestad}, which shows very clearly that $w < -1$ is a real possibility. The opposition is apparently motivated by a nebulous feeling that $w < -1$ is somehow ruled out by some basic principle. Let us consider what that principle might be.

A true equation of state, expressing a generally valid functional relationship between pressure and density, can be used to compute the speed of sound; and there is a natural reluctance to accept the possibility that the speed of sound might exceed that of light. [Actually, some authors \cite{kn:mukhanov} are willing to accept this, but we shall not do so here.] But if the magnitude of the pressure can exceed the density of the dark energy, then it seems that we run just this risk. In response to this we take note of two points. The first is that the ``cosmological equation of state" is not a true equation of state: it only expresses the relationship between pressure and density in the cosmological context -- that is, in a situation which is perfectly isotropic. Thus, {\em spatial} variations in the pressure and density are ruled out by fiat, and so the speed of sound cannot be reliably computed in this way. For example, the speed of sound in quintessence is {\em always} precisely equal to the speed of light, no matter what $w$ may be. Secondly, as in the case of quintessence, there is no guarantee that $w$ is independent of time, and so the elementary computation of the speed of sound [in terms of ${dp}\over{d\rho}$] is not to be trusted in any case. Of course, this is not to say that caution in this regard should be thrown to the winds -- certainly, we must be very careful if we are indeed assuming that $w < -1$ is a {\em constant} -- but it does mean that $w < -1$ does not necessarily imply a violation of local causality.

A more sophisticated version of the causality objection is that a theorem of Hawking and Ellis [discussed in \cite{kn:carter}] states that if the stress-tensor $T$ satisfies the Dominant Energy Condition [which forbids the absolute value of the pressure to exceed the density] and if $T$ vanishes on a closed achronal set $\Sigma$, then $T$ must also vanish on the domain of dependence of $\Sigma$. Thus $w \geq -1$ does indeed keep us safe from invasions of our domain of dependence. [Notice however that the Dominant Energy Condition does not protect us from {\em all} unwelcome behaviour -- in particular, the G\"odel spacetime has positive pressure equal to its density, so it satisfies the DEC.] However, this theorem does not have a converse -- if it did, then Anti-deSitter space, beloved of string theorists, which violates the DEC, would also allow invasions of domains of dependence. That is not the case, as the Penrose diagram immediately shows. Hence, once again, while we should be cautious in allowing $w < -1$, we should not be over-cautious: this range is {\em not} forbidden by causality.

An altogether different reason for viewing $w < -1$ with distaste is that, quite often, it leads to a truly bizarre prediction as to the way the world ends: not with a Crunch but with a Smash. That is, the cosmological scale factor $a(t)$ tends to infinity {\em in a finite time}. Thus, the world comes to an end not through excessive contraction but rather through excessive expansion. It is hard to say precisely what this would mean in reality, when local inhomogeneities are taken into account; perhaps the Universe would shatter into small gravitationally bound domains with ``singular boundaries" [hence the term ``Big Smash"]. The principal objection to this, apart from its sheer {\em bizarrerie}, is that there is little hope of resolving such a singularity using quantum-mechanical effects. [On the other hand, the Big Smash is surely the most natural way, granted that the expansion is indeed accelerating, of bringing the Universe to an end in a finite time; and that might ultimately prove the simplest explanation of the ``coincidence problem" -- it is not very surprising that
we find ourselves close to the onset of acceleration if the Universe comes to an end soon thereafter.] In any case,however, the Big Smash only follows from $w < -1$ if we assume that $w$
is {\em constant}. An explicit example of an infinitely long-lived cosmology with non-constant $w < -1$ has been given in \cite{kn:mcinnes}[see below]; thus, once again, we see that it is {\em not} necessarily the case that $w < -1$ leads to anything unpleasant.

Another objection to $w < -1$ is that it puts an end to {\em all} of the familiar energy conditions of General Relativity. ``All known forms of energy satisfy....." runs the familiar incantation, followed by some stricture on the relative sizes of the pressure and the density. It is difficult to see the relevance of this philosophy to our present predicament. For surely if there is one property of dark energy on which all are agreed, it is that the dark energy is {\em not} a ``known form of energy". Just why it should be expected to obey any of the energy conditions is, therefore, not easy to understand. Most theorists are quite happy to accept Anti-deSitter space, which violates the Weak and Dominant Energy Conditions, preserving only the Strong and the Null. The SEC is satisfied by ``all known forms of energy" -- but not by dark energy. As for the NEC: a cynic might be moved to suggest that it enjoys its current vogue primarily because one likes to have something agreeable to say about AdS. It is safe to predict that the NEC will be abandoned, if need be, without unduly incommoding many theorists.

\addtocounter{section}{2}
\section*{2. The View From String Theory}
What does string/M theory have to say about this matter, however? In such a theory, energy conditions must not be postulated independently: if they are valid, they must follow from the basic principles of the theory. In general, we certainly cannot hope to derive the Dominant Energy Condition from string theory. For, as is well known, string theory works very well 
on Anti-deSitter space, which, as we have emphasised, strongly violates the DEC. However, one could certainly imagine that while string theory tolerates violations of the DEC in general, perhaps it enforces the DEC in the special context of cosmology. 

In order to investigate this, the author constructed \cite{kn:mcinnes} a DEC-violating cosmology which is asymptotically deSitter both to the future and the past. The metric is 
$$g(\gamma ,A,L) = -dt \otimes dt + A^2 {\cosh}^{\left({4\over \gamma}\right)}
	\left( {\gamma t \over 2L}\right) (d\theta_1 \otimes d\theta_1 + 
d\theta_2 \otimes d\theta_2 + d\theta_3 \otimes d\theta_3), $$
where $\gamma$ is a parameter which measures the extent of DEC-violation, where $3/L^2$ is the asymptotic cosmological constant, and where the spatial sections are cubic flat tori with minimum side length $\pi A$. Unfortunately, it is not known precisely how to do string theory in such necessarily non-supersymmetric backgrounds, but Strominger \cite{kn:strominger1} and many others have proposed to circumvent this difficulty by postulating the existence of a ``dS/CFT correspondence", which is to be analogous to its celebrated AdS counterpart. [There is already a voluminous literature on this theme; the reader is asked to consult the list of papers citing and cited by \cite{kn:strominger1}; the authors in those lists are asked to forgive my failure to cite them here.] Thus one tries to deduce at least the gross features of the spacetime from the behaviour of conformal field theories inhabiting conformal infinity -- in the infinite future and past. In particular, the {\em flow of time itself} is interpreted \cite{kn:strominger2} as the dS/CFT counterpart of a renormalization group flow in the boundary CFT. When one applies these ideas, at least naively, to the above spacetime, one seems to find that time flows {\em towards} $t = 0$, from both positive and negative values of $t$. That does not make much sense physically, and so one has a piece of evidence that, in the cosmological context, string theory does not approve of violations of the DEC. However, it must be said that this evidence is rather weak: it is beginning to appear that, if indeed there is a dS/CFT correspondence, then the conformal field theories at infinity behave in a very unconventional way \cite{kn:bala}. In particular, the renormalization group flow may not behave in the usual manner. Thus the possibility that the above spacetime still makes sense in the dS/CFT context cannot be altogether ruled out. Meanwhile there remains the very difficult task of relating the dS/CFT correspondence to string theory in a convincing manner.

Regrettably, therefore, we have to conclude that string theory does not yet tell us anything very definite about the possibility that $w < -1$. However, it is not inconceivable that further investigations of the dS/CFT correspondence will ultimately lead to a concrete prediction that  $w \geq -1$. On the other hand, Frampton \cite{kn:framp} has recently argued that ``stringy" matter actually {\em requires} $w < -1$. In either case, string theory will have made a prediction which can perhaps be put to the test in the relatively near future: for if indeed $w \neq -1$, this may well be easier to demonstrate if $w < -1$ than if $w > -1$.


\begin{thebibliography}{18}
\linespread{0.5}
\bibitem{kn:sievers}
J. L. Sievers, J. R. Bond, J. K. Cartwright, C. R. Contaldi, B. S. Mason, S. T. Myers, S. Padin, T. J. Pearson, U.-L. Pen, D. Pogosyan, S. Prunet, A. C. S. Readhead, M. C. Shepherd, P. S. Udomprasert, L. Bronfman, W. L. Holzapfel, J. May, Cosmological Parameters from Cosmic Background Imager Observations and Comparisons with BOOMERANG, DASI, and MAXIMA,
astro-ph/0205387.
\bibitem{kn:caldwell}
R.R. Caldwell, A Phantom Menace?, Phys.Lett. B545 (2002) 23, arXiv:astro-ph/9908168
\bibitem{kn:hannestad}
S. Hannestad, E. Mortsell, Probing the dark side: Constraints on the dark energy equation of state from CMB, large scale structure and Type Ia supernovae, arXiv:astro-ph/0205096 
\bibitem{kn:mukhanov}
C. Armendariz-Picon, V. Mukhanov, P. J. Steinhardt, Essentials of k-essence, Phys.Rev. D63 (2001) 103510, arXiv:astro-ph/0006373 
\bibitem{kn:carter}
B. Carter, Energy dominance and the Hawking Ellis vacuum conservation theorem,
gr-qc/0205010.
\bibitem{kn:mcinnes}
B.McInnes, The dS/CFT Correspondence and the Big Smash, JHEP 0208 (2002) 029, 
arXiv:hep-th/0112066 
\bibitem{kn:strominger1}
A. Strominger, The dS/CFT Correspondence, JHEP 0110 (2001) 034, arXiv:hep-th/0106099
\bibitem{kn:strominger2}
A. Strominger, Inflation and the dS/CFT Correspondence, JHEP 0111 (2001) 049, 
arXiv:hep-th/0110087
\bibitem{kn:bala}
V. Balasubramanian, J. de Boer, D. Minic, Exploring de Sitter Space and Holography, 
arXiv:hep-th/0207245 
\bibitem{kn:framp}
P.H. Frampton, How to Test Stringy Dark Energy, arXiv:astro-ph/0209037 





\end{thebibliography}
\end{document}